\documentclass[twocolumn]{aastex63}
  \usepackage{graphicx}
 \shorttitle{LMC Gamma Ray power spectrum}
\shortauthors{Besserglik and Goldman}
\begin{document}
    
\title{ The power spectrum   and structure function of the Gamma Ray emission from the Large Magellanic Cloud }

\author{ Daniel Besserglik}  
 \affil{ Department of Astrophysics, Tel Aviv\   University,   Tel Aviv, Israel} 
 \author{Itzhak Goldman}
  \affil {Physics Department, Afeka Academic Engineering College,  Tel Aviv, Israel}
  \affil{ Department of Astrophysics, Tel Aviv\   University,   Tel Aviv, Israel}
 
\begin{abstract}
The Fermi-LAT observational data  of the diffuse  $\gamma$ ray emission from the Large Magellanic Cloud (LMC), were examined to test for the existence of underlying long range correlations.  A statistical test applied to the data indicated that the probability that data are random is $\sim 10^{-99}$. Thus we proceeded and have  
  used the counts-number data to compute  2D spatial autocorrelation,   power spectrum,  and structure function.

The most important result of the present study is a clear indication for large scale spatial underlying correlations. This is evident in {\bf all} the functions mentioned above. The 2D power spectrum has a logarithmic slope of -3 on large spatial scales and a logarithmic slope of -4 on small spatial scales. The structure function  has  logarithmic slopes equaling 1 and 2 for the large and small scales respectively. The logarithmic slopes of the structure function and the power spectrum are consistent.  

A plausible interpretation of these results is   the existence of a large scale  {\it compressible turbulence} with a 3D logarithmic slope of  -4  extending over the entire extent of the LMC. This   may reflect the  fact  that the  $\gamma$ Ray emission is in  star forming regions, where jets and shocks are abundant. 
  Both the power spectrum and structure function exhibit   steeper logarithmic  slopes for smaller spatial scales. This is interpreted as an indication that the turbulent region    has an effective depth of about 1.5 kpc.  

\end{abstract}


 \section{Introduction}
The Large Magellanic Cloud (LMC) is a satellite galaxy of the Milky Way galaxy. At a distance of $50$ kpc it is close enough to be studied with scrutiny. Indeed detailed observations with different wavelengths were carried out. Interestingly, even the first supernova neutrinos were detected from SN1987A,  in the LMC   \citep{Hirata+1987,Bionta+1987} .

\citet{Spicker+Feitzinger88} analyzed the HI data obtained by \citet{Rohlfs+84} whose spatial resolution was $\sim 200$~pc. They have used the data to calculate the autocorrelation structure function of the  emission weighted  
 velocity field.  They obtained a structure function compatible with turbulence on scales up to 1.5~kpc, which is steeper than the  structure function of  
 Kolmogorov turbulence \citep{Kolmogorov1941}.

\citet{elmegreen+2001} used the the $H_I$ emission intensity data   of the LMC,  obtained by 
 \citet{Kim+98}, 
to compute  the spatial power spectrum. The spatial resolution was  about 20~pc.
They derived a power-law that covered  2 decades of spatial scales in the range of $(20 \div 2000)$~pc  and  seemed consistent  with the Kolmogorov turbulence spectrum \citep{Kolmogorov1941}.   The power spectra showed a steepening at a scale of $(100\div 200)$~pc, that was interpreted as the $HI$ disk width. The observations were interpreted as indicating a density turbulence in the ISM of the LMC.  

\citet{block+10} analyzed the LMC infrared  Spitzer data  \citep{meixner+2006} and obtained spatial power spectra  spanning 3 orders of magnitudes  $(7 pc\div 7 kpc)$ extending over the entire size of the LMC. Here too a steepening at $100\div 200pc$  was observed.

The results of  \citet{elmegreen+2001}   and    \citet{block+10} suggest the existence of large scale turbulence in the interstellar medium (ISM) of the LMC. In this work we set to find out whether the  Fermi-Lat $\gamma Ray $
observations indicate the existence of   large scale spatial correlations and if so what is their nature,

 In section 2 we address the observational data. In section 3 we present the  analysis and we compute the 2D correlation, power spectrum and structure function of the observational data. The interpretation and implications are discussed in section 4. In appendices A and B,we obtain theoretical 2D power spectrum and structure function of data that are the result of integration along the line of sight. These are used to interpret the observational power spectrum and structure function. 
 
\section{data} 
  We  use the LMC $\gamma$ ray    data  of the Fermi-Lat collaboration  \citep{ackermann+16}, accumulated over a total observing time span of about 73 months. The region of interest (ROI) of the data covered an angular area of $10^o \times 10^o$; which considering the inclination    
 corresponds to    about $9$ by $9$ kpc. The  data  given  in a FiTS file\footnote{http://cdsarc.u-strasbg.fr/viz-bin/qcat?J} present   the observations    after subtraction of  a background model (see \citet{ackermann+16} for details). The pixels are   $0.1^o \times 0.1^o$ in size, namely $90\ pc  \times 90\ pc  $. The PSF is $0.2^o$, which equals two pixels,and the standard deviation at each position is 2 counts per pixel.  The data is centered around right ascension $\alpha=80. 894^0$   and declination $\delta = -69.756^o$. The counts are total counts   in the range of $0.2 -100$ GeV.

Figure 1. displays a 3D plot of the counts   per pixel in the ROI, as function of position. The counts peak at the large star forming region Dor30 (Tarntula Nebula).

   \begin{figure}[h!]
\includegraphics[scale=0.7]{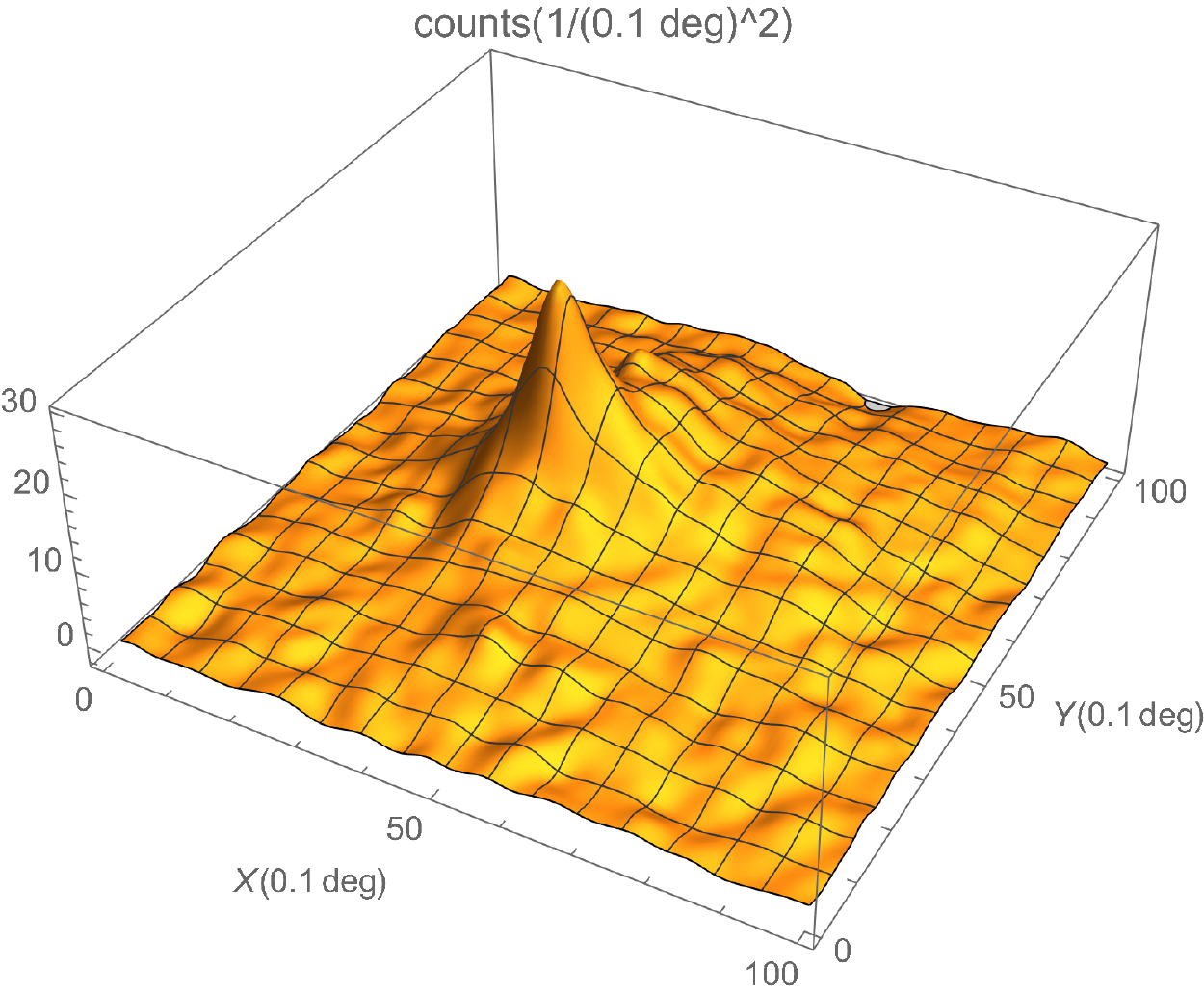} 
\caption{\label{fig:1}  3D plot of the counts per pixel in the ROI  as function of the X (right ascension)  and Y(declination) coordinates, in units of 0.1 deg.}
\end{figure}
  
\section{Analysis}

Before starting the data analysis we applied the {\it WolframMathemaica  AutoCorrelationTest} to the data.
This test estimates the probability of the hypothesis that  the {\bf data are random}. The result is
a probability $ \sim 10^{-99}$ implying that the data are {\it highly } auto-correlated. 
 
Figure 2. displays the 2D discrete normalized auto-correlation. The lags are in units of $0.1^o$.  It is seen that the correlation extends up to the edges of the ROI.
 
 In order to
study the nature of this long range spatial correlation 
  we apply two  analytical tools: 2D power spectrum and  2D structure function.
  The power spectrum is especially informative on the smaller spatial scales while  structure function complements it by a  better covering of the large spatial scales. The structure function has advantage  over the power spectrum in treating data at the maps edges (see e.g. \citet{Nestingen-Palm+2017}).

\vskip 1 cm
 \begin{figure}   
\includegraphics[scale=0.6]{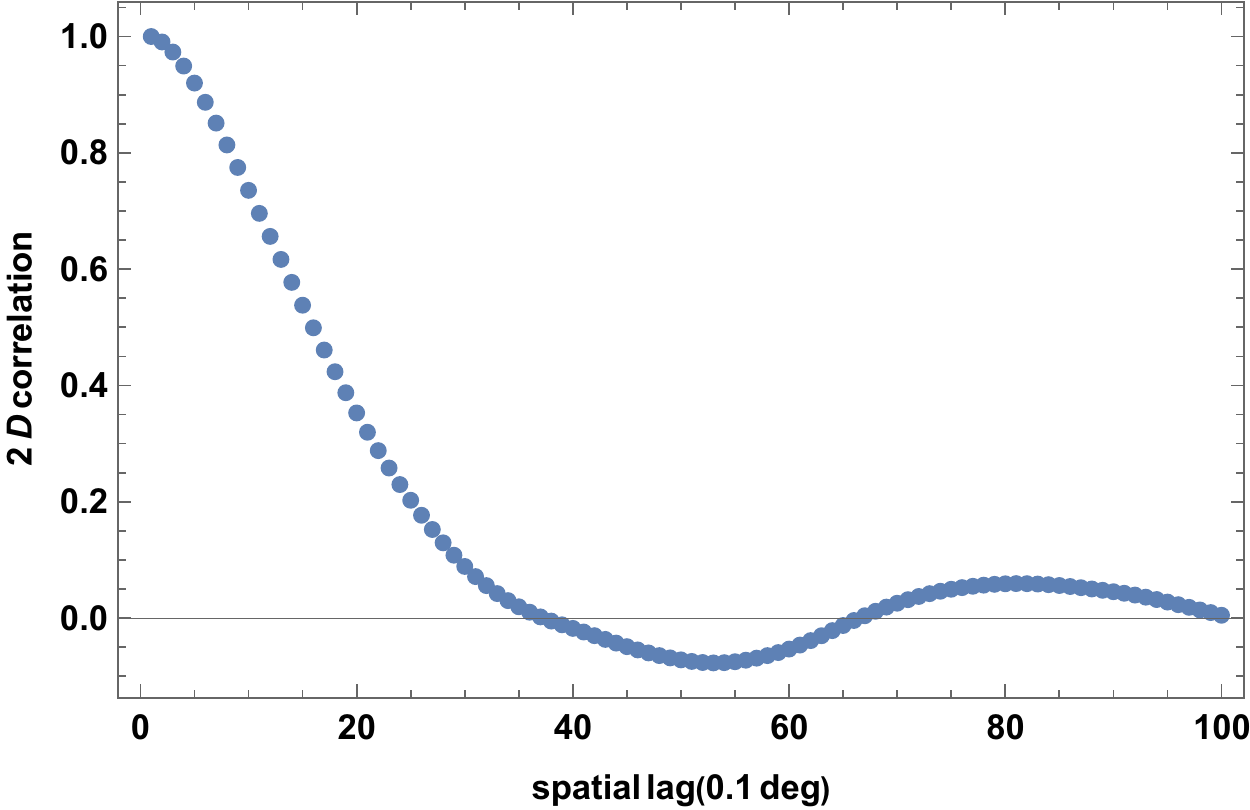} 
\caption{\label{fig:2} 2D normalized spatial auto-correlation  as function of spatial lag, in units of 0.1 deg}
\end{figure}

\subsection{2D power spectrum}
We computed the 2D power as function of the 2D wavenumber $k= (k_x^2 +k_y^2)^{1/2}$ by averaging over combinations of $k_x$ and $k_y$ that yield a given $k$.
The discrete FFT was used to compute the Fourier transform.

 The 2D power spectrum as function of the dimensionless wave-number $k$ is plotted in figure 3. The units
of the power spectrum are $deg^{-4}$, The dimensionless wave number $k$
 is defined as $ k= \frac{100}{r}$  
 with $r$ being the dimensionless spatial lag in units of $0.1 \ deg$.
 
 The  observational power spectrum exhibits  a logarithmic slop  of $-3$ for the large scales (small $k$) and $-4$ for small scales (large $k$). The logarithmic slope changes at $k\sim 3.5$, which corresponds to a spatial transition scale  $r_t \sim 2.57 \ kpc$.  At wave numbers $\gtrsim 17$ corresponding   to a scale $\lesssim 0. 5$~ kpc,  the power spectrum 
steepens considerably. 

 The error bars were computed by generating simulated data sets   
 that are randomly displaced from the observational values  and are  within twice the observational standard deviation. For each set the power spectrum was computed, and the standard deviations of the power spectrum were obtained. The error bars in the figures are $2\sigma$.

\vskip 1 cm
 \begin{figure}
.\label{ps}  
\includegraphics[scale=0.65]{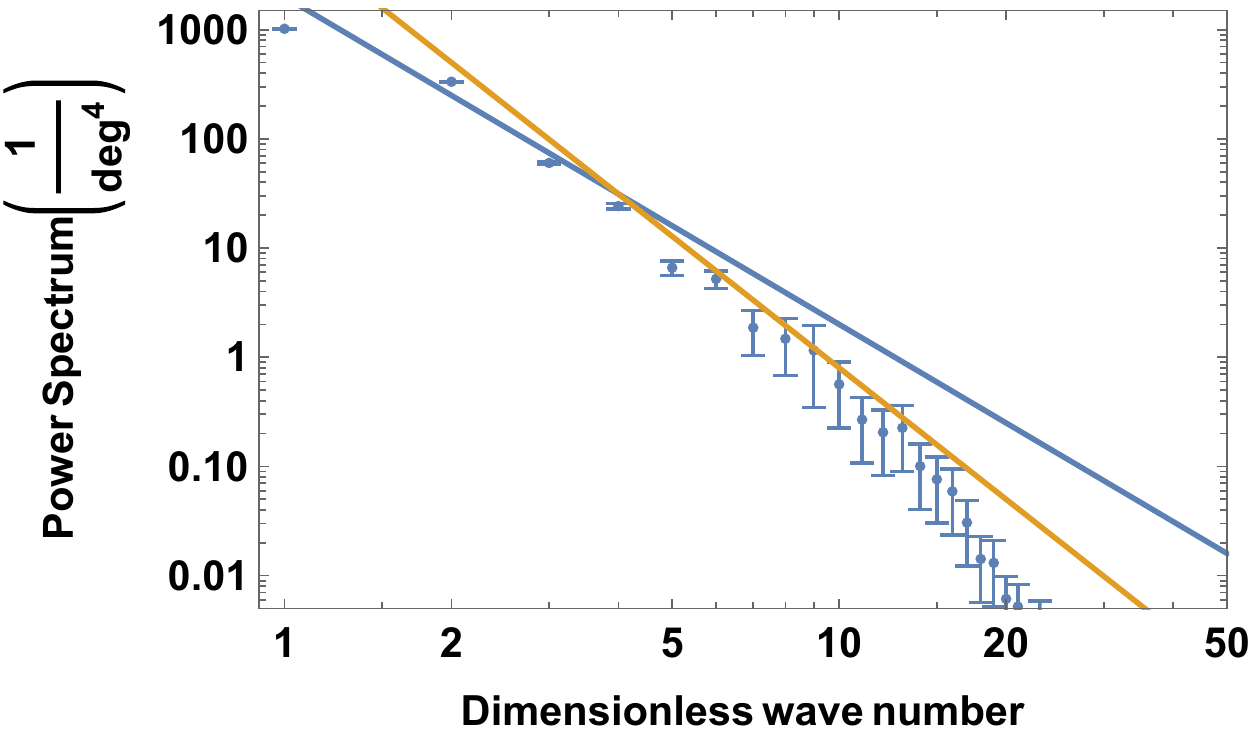} 
\caption{ 2D spatial power spectrum. The lines have logarithmic slopes of -3 (blue line) and -4 ( orange line) }
\end{figure}
 
\subsection{2D structure  spectrum}

The 2D  structure function $S_2(r)$ of a  2D quantity $f( x, y)$ is  

\begin{eqnarray}
\label{struct}
\hskip -0.7 cm  S_2(r)=<\left(f(x', y') -f( x'+x, y'+y)\right)^2>\\
 ~= 2 \bigg(C_2 (0) -C_2(r)\bigg) \ \ ; \ \ r=\sqrt{x^2 + y^2} \nonumber
\end{eqnarray}

where $r$ is the 2D lag between positions. The angular brackets are ensemble average which, by using the ergodic principle, can be replaced by space-average; in this case over the 2D x-y plane.

Figure 4. shows the 2D structure function computed from the data.  The structure function shows   a transition from a logarithmic slope of 2 for small spatial  lags  to a slope of 1 for large spatial lags.

 \begin{figure}[h!]
.\label{sf}  
\includegraphics[scale=0.65]{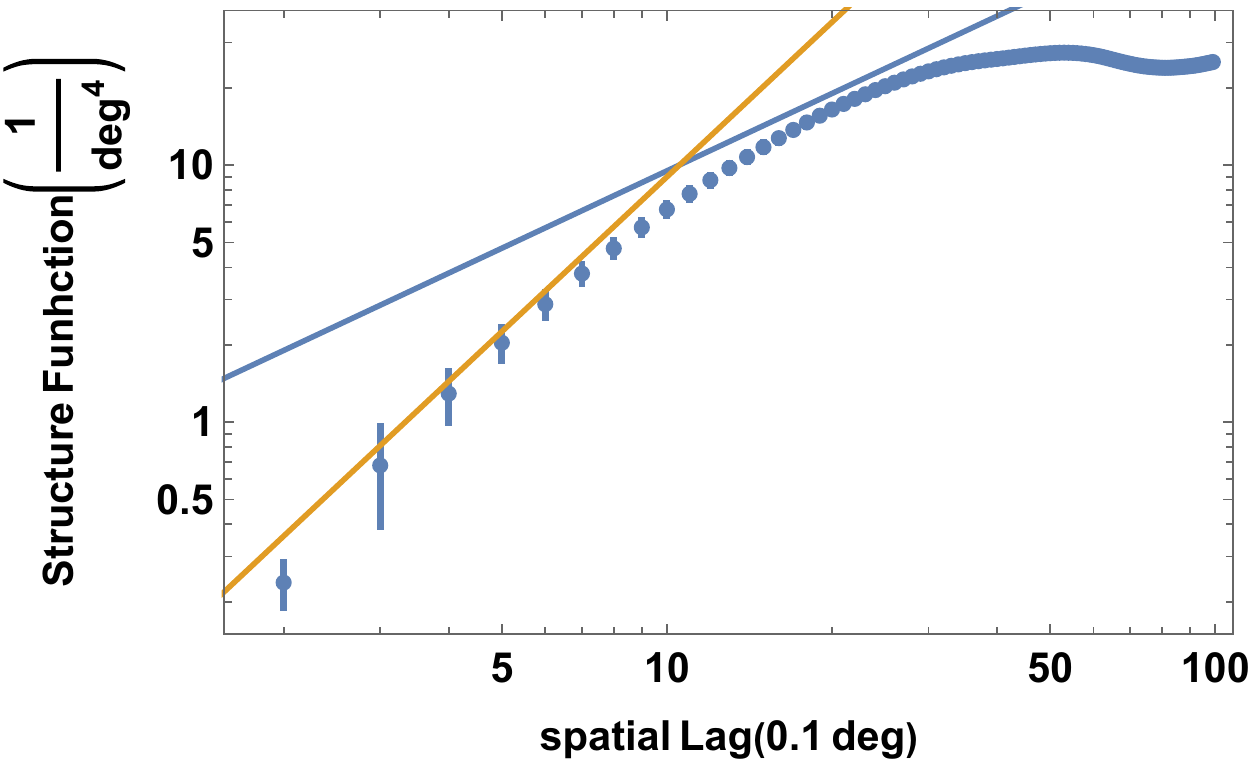} 
\caption{ 2D spatial structure function. The lines have logarithmic slopes of 1(blue line), and 2 ( orange line) }   
 \end{figure}
 
  Note that the structure function provides a complimentary description to that of the power spectrum. It provides more points on the larger spatial scales while the power spectrum provides more points on the 
smaller spatial scales.

 The observational transition lag is $r_t\sim 1.1^o $ corresponding to  $\sim 0.99$ kpc.

\section{discussion}

\subsection{The nature of the  turbulence}

The logarithmic slopes of the observational 2D power spectrum and of the 2D structure function are those expected for a compressible supersonic  turbulence.  Such turbulent power spectra were observed in HI  intensity maps in the Milky Way (MW) galaxy \citep{Green93}and in the SMC \citep{Stanimirovic+99}. This power spectrum has been observed  in molecular clouds \citep{Leung+82}, in the   HII region Sharpless~142 \citep{Roy+Joncas85}, and in numerical simulations \citep{Passot+88}.

 The 3D   power spectrum is  proportional $k{-4}$ with $k$ the absolute value of the 3D wavenumber (and equivalently a 1D power spectrum with logarithmic slope of -2). This is steeper than the Kolmogorov spectrum, which describes subsonic incompressible turbulence
with a 1D logarithmic slope of $-5/3$ and a 3D logarithmic slope of $-11/3$. 

The steeper slope signals that (unlike in the Kolmogorov spectrum)
the rate of energy transferred in the turbulence cascade is not constant but  decreases with increasing wavenumber. This is indeed expected in a compressible turbulence since part of the energy at a given wavenumber in the cascade, is diverted to compression of the gas.
The existence of this turbulence is in line with observations by  \citet{Castro+2018} of supersonic $H\alpha$ velocity dispersions of $\sim 40\div30\ km s^{-1}$   in star forming regions in the LMC.   
  \subsection{The depth of the emitting region}
  
The observed $\gamma$ ray photons  originate from different depths along the line of sight.  
 Several authors addressed the issue of power spectra of quantities which  are the result of integration along the line-of-sight  \citep{{Stutzki+98}, {Goldman2000}, {Lazarian+Pogosyan2000},  {Miville+2003a}}. They concluded that when the lateral spatial scale is smaller than the depth of the layer, the logarithmic slope of the power spectrum 	 steepens exactly by $-1$ compared to its value when the lateral scale is large compared to the depth. This behavior was indeed  found in   as observational power spectra  of Galactic and extra Galactic turbulence ( e.g. 
\citet{elmegreen+2001}, \citet{Miville+2003b} ) and in solar photospheric turbulence \citep{Abramenko+Yurchyshyn2020}.

In Appendix A  and Appendix B,  we 
obtain  theoretical power spectra and structure function that are the result of integration along the line of sight, For the compressible turbulence we find that the transition from a $-3$ to a $- 4$  logarithmic slope occurs at a transition spatial lag $r_t= 1.67 D$ where $D$ is the effective depth from which the the emission originates. The observational value of $r_t =2.57 $kpc implies a depth $D=1.54$ kpc.

For the structure function we obtained that the transition lag from a logarithmic  slope of 1 to a logarithmic slope of $2$ occurs when $r_{SF}= 0.63 D$. Combining it with observations value obtained in the previous section $r_{SF}= 0.99$~kpc, implies a depth $D= 1.57$~kpc, in excellent agreement with the value implied by the observational power spectrum.  

The effective depth inferred from the power spectrum and structure function is an {\it actual} depth. of the turbulent layer 
  because the mean free path of $\gamma$ ray photons turns out to be much larger, as seen below. We note that this depth is an order of magnitude larger than  the HI depth obtained by 
 \citet{elmegreen+2001}. The large depth traces the depth of star forming regions where the cosmic rays are produced. In this context it is of interest  to note  that the stellar depth of the LMC,  is comparable to   its lateral dimensions   \citep{Jacyszyn+2017,  Subramanian2009, Subramanian2010}. As the lateral dimensions of the star formation regions (notably Dor 30) are
 $\sim 2$~kpc, the derived depth here makes sense.

\subsubsection{The $\gamma$ ray optical depth of the LMC}
In order to estimate the optical depth of the $\gamma $ ray photons in the LMC, we use the Klein Nishima 
cross-section, For photon energies much larger than the rest mass energy of the electron we use \citep{Neronov2017}.
 \begin{equation}
  \sigma_{KN}=\frac{3}{8}\sigma_T\frac{ \ln(2 x)}{x}\ \ \ ,   \ \ \ x=\frac{E_{\gamma}}{m_e c^2} 
  \end{equation}
with $\sigma_T=  6.65 \times 10^{-25}\ cm^2$ ,  a typical number density  (in  star forming  regions) of $10^3\ {\rm cm^{-3}}$, even the lowest energy photons have 
 mean free path  
 \begin{equation}
 l_{\gamma} \sim 80kpc
 \end{equation}
  
 This is an  order  of magnitude  larger than the size of the LMC.
  
 \subsection{Implication of the observational power spectrum on the cosmic rays spatial distribution}  
 The most likely mechanism for the production of the $\gamma $ rays,  adopted  also by \citet{ackermann+16},  is that of 
       decay  of pions    created by energetic cosmic ray (CR) protons scattering off the protons in the LMC  interstellar medium. The CR are produced inside the LMC in the star forming regions, notably in   30 Dorados, and confined by the LMC magnetic field. The latter has been investigated by \citet{Gaensler+2005} and   \citet{Mao+2012}who found a mean field along the line of sight of $\sim 1 \mu G$ and a disordered field with a coherence length of about $100$ pc and strength  of $\sim 10 \mu G$. It is conceivable that the weaker mean field is a result of a random walk of the disordered small scale  stronger field  \citep{Han2017}.
       
The cosmic rays are thought to diffuse along the magnetic field lines. The diffusion coefficient is large so that the cosmic ray population tends to be homogeneously spread   \citep{Grenier+2015, Krumholz+2020}. 
     
The local $\gamma$ ray  emissivity is proportional to the product of the number density of the CR and that of the interstellar medium protons.  The protons (hydrogen atoms and ions) are those who manifest turbulent fluctuations in velocity and number density. This turbulence is supersonic and is expected to have a 3D power spectrum with logarithmic derivative of -4.
The fact that the observational power spectrum is identical to this power spectrum implies that the CR population is indeed  homogeneous.

 Moreover, since the $\gamma$ ray emission is proportional to the proton number density, it is clear that the $\gamma$ ray  production comes  practically from  the star forming regions where the number density is more than 3 orders of magnitude larger than that in other  parts of the   ISM.

\subsection{Turbulence dissipation  scale}

As noted in section 3, for $k\gtrsim 17$, corresponding to a spatial scale  which is  $\lesssim  0.5 $~kpc,  the power spectrum steepens quite drastically. This  may mark  the scale below which the microscopic molecular viscosity dissipates the turbulent energy. In what follows, we wish to estimate the expected value of the dissipation scale and compare it with the observational one.

The dissipation scale is defined as  the scale below which the microscopic viscosity is larger than the turbulent viscosity, implying that the rate of energy dissipation by the microscopic viscosity exceeds the rate of energy cascaded by the turbulence. Denoting the transition scale by   $l_d$, and 
 the turbulent kinematic viscosity   on this scale,   by $\nu_t(l_d)$ one has for    the 1D  power spectrum which is proportional to $k^{-2}$

\begin{equation}
\nu_t(l_d) \sim \frac{1}{3} V_t(l_d) l_d \ ; V_t(l_d)= V_t(l_0) (\frac{l_d}{l_0})^{0.5}
\end{equation}
 
with $V_t(l_d)$ the turbulent velocity of the dissipation scale, $V_t(l_0)$ is the turbulent velocity on the largest scale  $l_0= 9$~kpc. 

The microscopic kinematic viscosity  $\nu_m$ is
$$\nu_m=\frac{1}{3}{c_s l_f}$$
 where $c_s$  is the sound speed and $l_f$ is the effective mean free path for atoms or ions  collisions.
 
The effective mean free path for the ionized H atoms is the coherence length of the fluctuating magnetic field  which is $\sim 100 $~pc \citep{Gaensler+2005}. The neutral H atoms are coupled to the ionized H atoms on a much smaller scale due to mutual scattering with crossection of $\sim 10^{-16} cm^2$.  
 
The  two viscosities are equal on the dissipation scale
 \begin{eqnarray}
  l_d= 
 450\ {\rm pc} \left (\frac{c_s} {V_t (l_0) }  \right)^{2/3} \left(  \frac{l_f}{100\ {\rm pc}} \right)^{2/3} 
 \end{eqnarray}
 
 This value is indeed consistent with the value suggested by the observational power spectrum.

  \section{Summary and conclusions}
 \begin{itemize}
\item 
The main result of the present work is revealing that the LMC $\gamma$ ray intensity exhibits    spatial correlations over scales  comparable to the size of this galaxy. These correlations manifest via the observational
2D power spectrum and structure function of the   $\gamma$ ray intensity. The power spectrum and the structure function are those of compressible supersonic turbulence.  The emerging   scenario is that of  a turbulent ISM and  cosmic ray protons which  are distributed rather homogeneously. This is consistent with models of cosmic rays diffusion along field lines that suggest such homogeneous distribution..
\item
The logarithmic slope of the power spectrum changes from $-3$ on large scales to $-4$ on small scales.
 The logarithmic slope of the structure function  changes from $1$ on large scales to $2$ on small scales.
This is indeed expected for data which is an integral over the line of sight.  Theoretical power spectrum and structure function,  detailed  in   appendix A and appendix B,,
were used to infer the depth from the observational power spectrum and structure function. 
 The  resulting  depth of the emitting region,  from both,   is $\sim 1.5$ kpc. 
 The large depth reflects the depth of star forming regions where the cosmic rays are produced.
 \item
The large scale of the turbulence requires a generating mechanism  which acts on such a global scale. Following \citet{Goldman2000} we suggest  that the source generating the turbulence is the tidal interaction with the SMC. The last close passage of the two Magellanic clouds occurred about $200$ MYR ago \citep{Gardiner+Noguchi1996, Yoshizawa+Noguchi2003} . 
 
 Assuming supersonic  turbulent velocity of $\sim 30 km s^{-1}$ and largest spatial scale of $\sim 9$ kpc    the decay time is $\sim  300$~Myr.  Thus the turbulence has not decayed yet.
  On more local scales  of  $\sim 1$~kpc   there is energy injection from supernovae and jets in the star forming regions.
\end{itemize}
 
\section*{Acknowledgment} 
  We thank Shmuel Nussinov for suggestions and comments. Itzhak Goldman thanks the Afeka College Research Authority for support.
 
 \appendix 

  \section{2D power spectrum of data integrated along the line-of-sight}

  We are interested in the power spectrum of the counts per unit area in the plane of the sky, $n(\vec r)$  which  is an integral along the line-of -sight  $z$ of the counts per unit volume  $f( \vec r, z)$. Here $\vec r= ( x, y)$ is a position in in the plane of the sky.
  \begin{equation}
 n(\vec r)  = \int_0 ^{D } f( \vec r, z)dz  
   \end{equation}   
with $D$ denoting the depth of the turbulence along the line of sight. 
   
The 2-dimensional  power spectrum of $n(\vec r)$ which depends also on  $D$ is
  
  \begin{equation}
P_2(\vec k, D) = \int e^{- i \vec k \cdot \vec r} C_2\left( n(\vec r) \right)    d^2\vec r
  \end{equation}
   with  $C_2 ( \vec r)  $ being the 2D  2-point autocorrelation of the fluctuating 
   $n(\vec r)$.  
   
  \begin{eqnarray}
       C_2( \vec r )= <n( \vec r') n(\vec r' + \vec r) >=
   \int_{0} ^{D }\int_{0} ^{D}< f(\vec {r}\ ' , z')(f( \vec r' + \vec r ,   z) >    dzdz'    =\\
   \nonumber
  \int_{0} ^{D }\int_{0} ^{D}C_3 (\vec r, z-z')    dzdz' 
   =\int_{0} ^{D }\int_{0} ^{D} P_3 ( \vec k, k_z)
  e^{-i \vec k\cdot \vec r -i  k_z ( z-z')}d^2 \vec k dzdz'
   \end{eqnarray}
Here $C_3 (\vec r, z-z')$ and   $P_3 ( \vec k, k_z)$ are the 3D autocorrelation and power spectrum, respectively.

From equation (A2) we identify the 2D power spectrum.
\begin{eqnarray}
P_2( \vec k) = \int_{-\infty}^{\infty} \int_{0} ^{D }\int_{0}^D P_3 ( \vec k, k_z ) e^{i  k_z ( z-z')}dzdz'dk_z \propto \int_0^{\infty} P_3 ( \vec k, k_z ) \left(\frac{\sin( k_z D/2)}{k_z D/2} \right)^2 dk_z 
\end{eqnarray}   
   
   When the 3D power spectrum is  a power law    and is a function of $k= |{\vec k}|$ 
   \begin{equation}
    P_3 ( \vec k, k_z ) \propto ( k^2 + k_z)^{-(m+2)}
    \end{equation}
 the 2D power spectrum becomes
    
 \begin{equation} 
      P_2( \vec k) \propto \int_0^{\infty} ( k^2 + k_z)^{-(m+2)} \left(\frac{\sin( k_z D/2)}{k_z D/2} \right)^2 dk_z  
  \end{equation}    
 
For the present case, $m=2$, and using the dimensionless variable  $\eta=k D/2$  one gets an analytic solution
    
     \begin{equation}
    P_2(\eta) \ \propto  \eta^{-4}   \left(  \cosh  \eta -\sinh \eta \right) \left(3 \eta^{-1}
 (\eta\cosh  \eta -\sinh  \eta)+\sinh  \eta \right)   
 \end{equation} 
   
    \begin{figure}[h!]
.\label{psth}  
\includegraphics[scale=0.7]{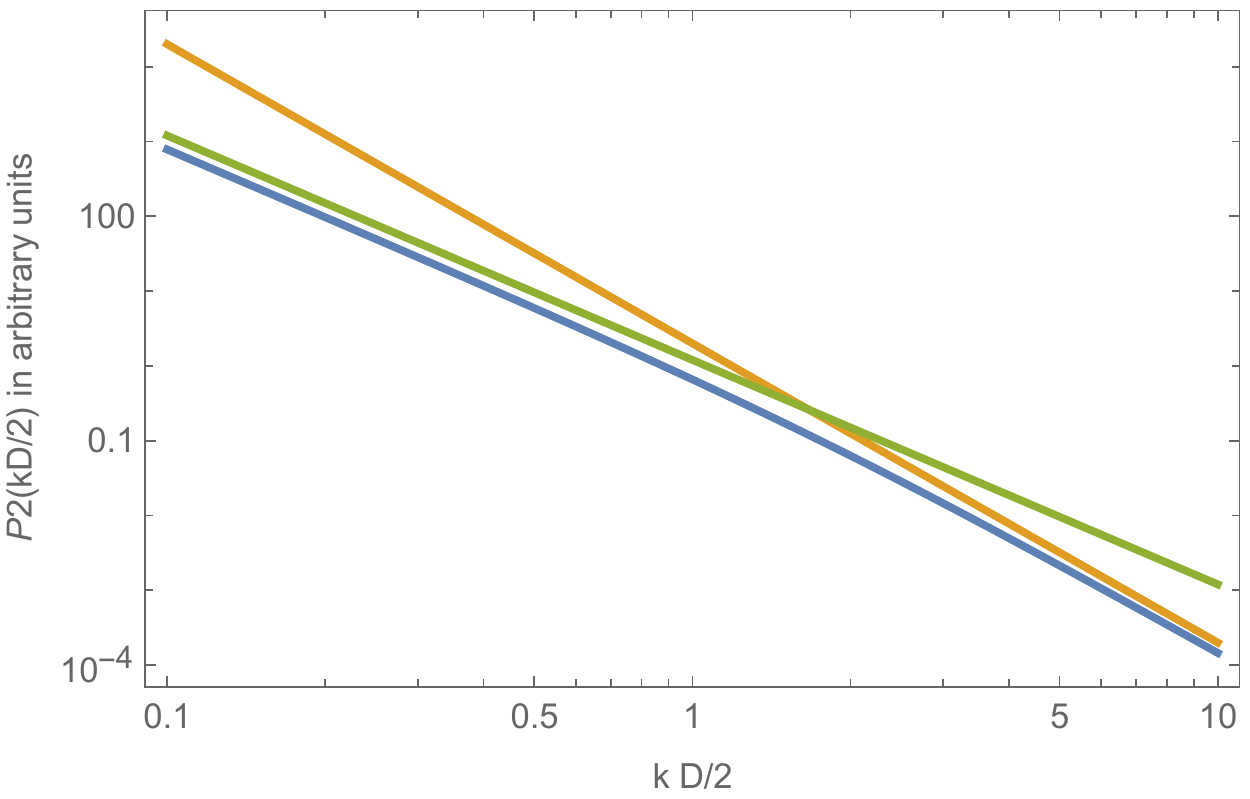} 
\caption{ 2D theoretical spatial power spectrum for m=2. The lines have logarithmic slopes of -3 (green line) and -4 (orange line) }
\end{figure}

 Figure 5. displays   $P_2(kD/2)$. It is seen that for $\eta >>1$ the logarithmic slope is -4 while in the limit 
$\eta <<1$ the logarithmic slope is -3. A tangent to the curve with a logarithmic slope of $-3.5$ was used (not shown here)  to define the transition value of $\eta$ : $\eta_t = k_t D/2= 1.88$ so that $D = 3.76/ k_t= 0.6 r_t$. Here, $k_t$ is the transition wave number and  $r_t= 2\pi /k_t $ is the transition spatial lag.

  \section{The 2D structure function of data integrated along the line-of-sight }

 The 2D correlation  is obtained from the 2D power spectrum via
 
 \begin{equation} 
 C_2(\vec r)= \int_{-\infty}^{\infty} P_2( \vec k)e^{i\vec k\cdot \vec r}d^2 \vec k
  \end{equation}

Using equation (1) and taking note that in 
   the present case, $P_2(\vec k) $ and $C_2 (\vec r)$ are functions of the absolute values of $k$, and $r$, respectively, we get 
 
 \begin{eqnarray}
    S_2(  r)\propto \int_0^{\infty}\sin^2( kr/2) \int_0^{2\pi} P_2(   k)e^{i k r \cos\theta} d\theta   k dk \propto \int_0^{\infty}\sin^2( \eta r/D) \int_0^{ 2\pi} P_2(  \eta) e^{i \eta r/D \cos\theta} d\theta  \eta d\eta 
     \end{eqnarray} 
      performing the integration over $\theta$, the angle between $\vec k$ and $ \vec r$, yields
 \begin{eqnarray}
       S_2(  r)   \propto \int_0^{\infty}\sin^2( \eta r/D)   P_2(  \eta)     
 \pi(1 - J_0( \eta r/Dr)\  \eta\ d\eta \\
 \nonumber
   \propto  \int_0^{\infty}\sin^2( \eta r/D)  \eta^{-4}   \left(  \cosh  \eta -\sinh \eta \right) \left(3 \eta^{-1} 
 (\eta\cosh  \eta -\sinh  \eta)+\sinh  \eta \right)\eta d\eta    
     \end{eqnarray}  
 Figure 6. displays   $S_2(r/D)$. It is seen that for $r/D>>1$ the logarithmic slope is 1 while in the limit  
$r/D<<1$ the logarithmic slope is 2. A tangent to the curve with a logarithmic slope of $ 1.5$ was  used to define the transition value  $r_t/D= 0.63$  implying $D= 1.59 r_t$ with $r_t$ denoting the transition spatial scale.
 
    \begin{figure}[h!]
.\label{sfth}  
\includegraphics[scale=0.4]{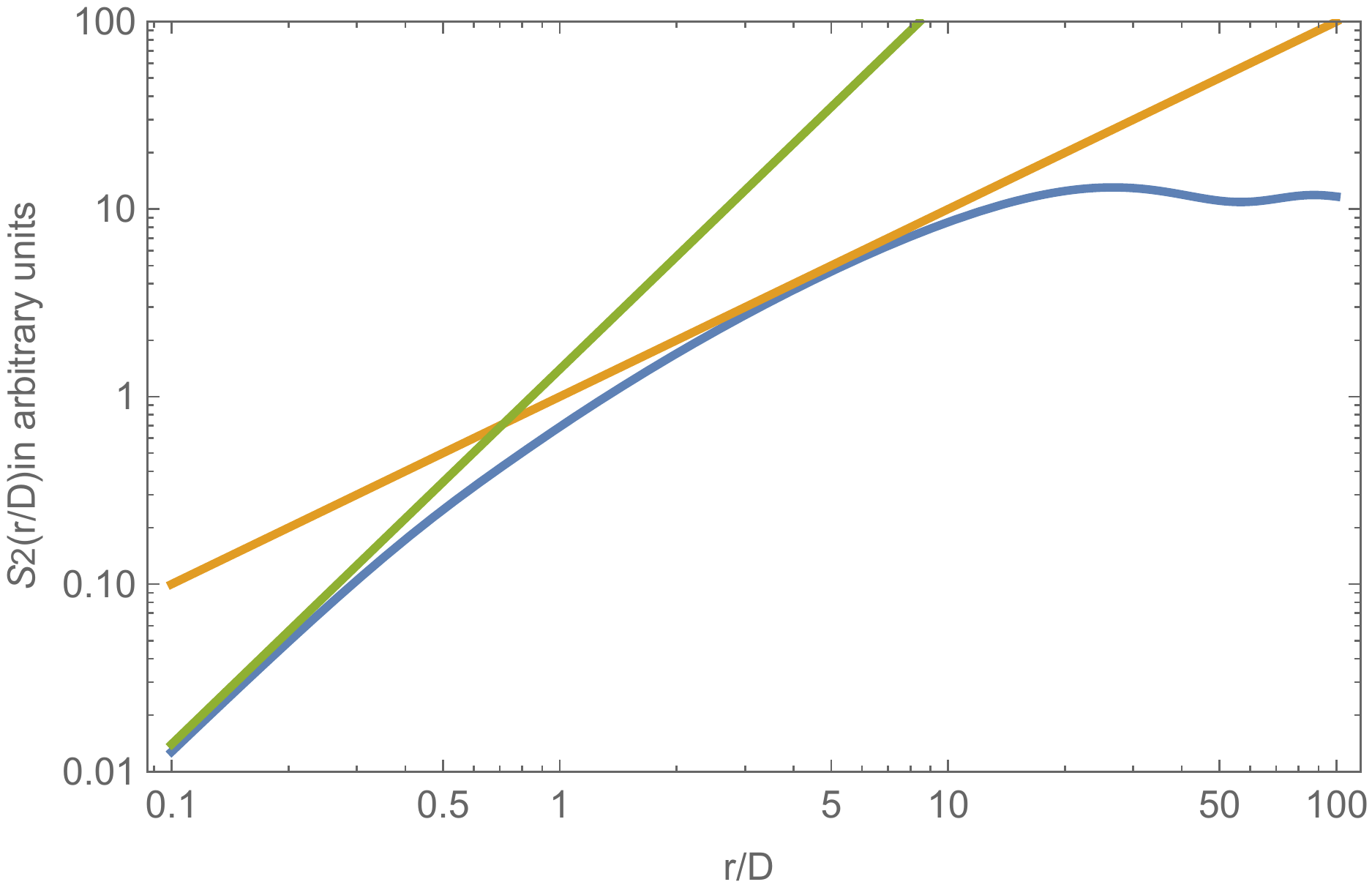} 
\caption{ 2D theoretical spatial structure function for m=2. The lines have logarithmic slopes of 1 (orange line ) and 2 (green line) }
\end{figure}

\end{document}